\documentclass[prb, aps, floatfix, twocolumn]{revtex4-1}
\usepackage{CJK}
\usepackage{pdfsync}
\usepackage{graphicx,bm}
\usepackage{subfigure} 
\usepackage{amssymb}
\usepackage{psfrag}
\usepackage{amsmath}
\usepackage{verbatim}
\usepackage{my_symbols}
\usepackage{amsfonts}
\usepackage{graphicx}

\newcommand{\HHeff}{\ensuremath{{\bf H}_{\rm eff}}} 
\newcommand{\Heff}{\ensuremath{H_{\rm eff}}} 
\newcommand{\II}{\ensuremath{{\bf I}}}

\begin{document}
\begin{CJK*}{UTF8}{gbsn} 

\title{Theory of magnon-driven spin Seebeck effect}

\author{Jiang Xiao (萧江)$^{1,2}$, Gerrit E. W. Bauer$^2$, Ken-ichi Uchida$^{3,4}$, Eiji
Saitoh$^{3,4,5}$, and Sadamichi Maekawa$^{3,6}$}


\affiliation{$^1$Department of Physics and State Key Laboratory of Surface Physics, Fudan
University, Shanghai 200433, China \\ $^2$Kavli Institute of NanoScience, Delft University of
Technology, 2628 CJ Delft, The Netherlands \\ $^3$Institute for Materials Research, Tohoku
University, Sendai 980-8557, Japan \\ $^4$Department of Applied Physics and
Physico-Informatics, Keio University, Yokohama 223-8522, Japan \\ $^5$PRESTO, Japan Science and
Technology Agency, Sanbancho, Tokyo 102-0075, Japan \\ $^6$CREST, Japan Science and Technology
Agency, Tokyo 100-0075, Japan}

\begin{abstract}

The spin Seebeck effect is a spin-motive force generated by a temperature gradient in a
ferromagnet that can be detected via normal metal contacts through the inverse spin Hall effect
[K. Uchida {\it et al.}, Nature {\bf 455}, 778-781 (2008)]. We explain this effect by spin
pumping at the contact that is proportional to the spin-mixing conductance of the interface,
the inverse of a temperature-dependent magnetic coherence volume, and the difference between
the magnon temperature in the ferromagnet and the electron temperature in the normal metal [D.
J. Sanders and D. Walton, Phys. Rev. B {\bf 15}, 1489 (1977)]. 

\end{abstract} 
\date{\today} \maketitle
\end{CJK*}


\section{Introduction}

The emerging field called {\it spin caloritronics} addresses charge and heat flow in
spin-polarized materials, structures, and devices. Most thermoelectric phenomena can depend on
spin, as discussed by many authors in Ref. \onlinecite{special_issue}. A recent and not yet
fully explained experiment \cite{uchida_observation_2008} is the spin anologue of the Seebeck
effect --- the spin Seebeck effect, in which a temperature gradient over a ferromagnet gives
rise to an inverse spin Hall voltage signal in an attached Pt electrode.

\begin{figure}[b]
	\includegraphics[width=0.9\columnwidth]{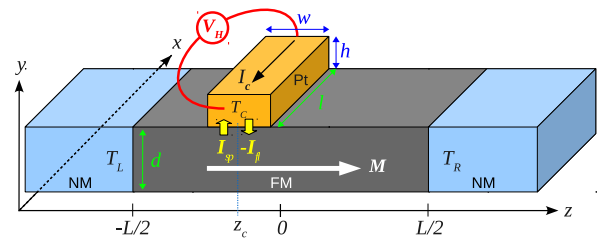}
	\caption{(Color online) A ferromagnet (F) with thickness $d$ and length $L$ and
	magnetization $\MM$ pointing in $\hzz$-direction connects two normal metal (N) contacts at
	temperature $T_{L/R}$ at the left and right ends. A Pt strip of dimension $l{\times} w {\times} h$ on
	top of F converts an injected spin current ($\II_s = \II_{sp} + \II_{fl}$) into an
	electrical current $I_c$ or Hall voltage $V_H$ by the inverse spin Hall effect.}
	\label{fig:NFNC}
\end{figure}

The Seebeck effect refers to the electrical current/voltage that is induced when a temperature
bias is applied across a conductor. By connecting two conductors with different Seebeck
coefficients electrically at one end at a certain temperature, a voltage can be be measured
between the other two ends when kept at a different temperature. The spin counterpart of such a
thermocouple is the spin current/accumulation that is induced by a temperature difference
applied across a ferromagnet, interpreting the two spin channels as the two ``conductors''. In
Uchida {\it et al.}'s experiment, \cite{uchida_observation_2008} a temperature bias is applied
over a strip of a ferromagnetic film. A thermally induced spin signal is measured by the
voltage induced by the inverse spin Hall effect (ISHE) \cite{valenzuela_direct_2006,
kimura_room-temperature_2007} in Pt contacts on top of the film in transverse direction (see
\Figure{fig:NFNC}). This Hall voltage is found to be approximately a linear function (possibly
a hyperbolic function with long decay length) of the position in longitudinal direction over a
length of several millimeters. This result has been puzzling, since spin-dependent length
scales are usually much smaller. The original explanation for this experiment has been based on
the thermally induced spin accumulation in terms of the spin thermocouple analogue mentioned
above. However, the spin flip scattering shot-circuits the spin channels, and at which spin
channels are shot-circuited, the signal should vanish on the scale of the spin-flip diffusion
length. \cite{hatami_2009}

In this paper, we propose an alternative mechanism in terms of spin pumping caused by the
difference between the magnon temperature in the ferromagnetic film and the electron
temperature (assumed equal to the phonon temperature) in the Pt contact. Such a temperature
difference can be generated by a temperature bias applied over the ferromagnetic film.
\cite{sanders_effect_1977}

This paper is organized as follows: Section \ref{sec:Isz} describes how a DC spin current is
pumped through a ferromagnet(F)\big|normal metal(N) interface by a difference between the
magnon temperature in F and electron temperature in N. In section \ref{sec:TmTp} we calculate
the magnon temperature profile in F under a temperature bias. In Section \ref{sec:VH} we
compute the thermally driven spin current as a function of the position of the normal metal
contact.

\section{Thermally driven spin pumping current across F\big|N interface}
\label{sec:Isz}

\begin{figure}[b]
	\includegraphics[width=0.7\columnwidth]{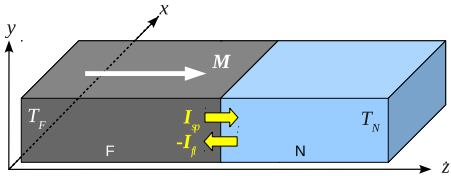}
	\caption{(Color online) F\big|N interface with 1) spin pumping current $\II_{sp}$ driven
	by the thermal activation of the magnetization in F at temperature $T_F$, and 2)
	fluctuating spin current $\II_{fl}$ driven by the thermal activation of the electron spins
	in N at temperature $T_N$.}
	\label{fig:FN}
\end{figure}

In this Section we derive expressions for the spin current flowing through an F\big|N interface
with a temperature difference as shown in \Figure{fig:FN}, starting with the macrospin
approximation in Subsection II.A and considering finite magnon dispersion in Subsection II.B.

Since the relaxation times in the spin, phonon, and electron subsystems are much shorter than
the spin-lattice relaxation time, \cite{kittel_relaxation_1953, demokritov_bose-einstein_2006}
the reservoirs become thermalized internally before they equilibrate with each other.
Therefore, we may assume that the phonon (p), conduction electron (e), and magnon (m)
subsystems can be described by their local temperatures: $T_F^{p, e, m}$ in F, and $T_N^{p, e}$
in N. \cite{beaurepaire_ultrafast_1996} We furthermore assume that the electron-phonon
interaction is strong enough such that locally $T_F^p = T_F^e {\equiv} T_F$ and $T_N^p = T_N^e {\equiv}
T_N$. However, the magnon temperature may deviate: $T_F^m {\neq} T_F$. This is illustrated below
by the extreme case of the macrospin model, in which there is only one constant magnetic
temperature, whereas the electron and phonon temperatures linearly interpolate between the
reservoir temperatures $T_L$ and $T_R$. The difference between magnon and electron/phonon
temperature therefore changes sign in the center of the sample. When considering ferromagnetic
insulators, the conduction electron subsystem in F becomes irrelevant.

\subsection{F\big|N contact} \label{sub:nano}

First, let us consider a structure such as shown in \Figure{fig:FN}, in which the magnetization
is a single domain and can be regarded as a macrospin ${\bf M} = M_sV\mm$, where $\mm$ is the
unit vector parallel to the magnetization. We will derive in Subsection II.B the criteria for
the macrospin regime. We assume uniaxial anisotropy
along $\hzz$, $M_s$ is the saturation magnetization and $V$ is the total F volume. The
macrospin assumption will be relaxed below, but serves to illustrate the basic physics.

At finite temperature the magnetization order parameter in F is thermally activated, {\it i.e.}
$\dm \ne 0$. When we assume N to be an ideal reservoir, a spin current noise $\II_{sp}$ is
emitted into N due to spin pumping according to: \cite{tserkovnyak_enhanced_2002}
\begin{equation}
	\II_{sp}(t) = {{\hbar}\ov 4{\pi}}\midb{g_r ~\mm(t){\times}\dmm(t) + g_i~\dmm(t)},
\label{eqn:Isp}
\end{equation}
where $g_r$ and $g_i$ are the real and imaginary part of the spin-mixing conductance $g_{\rm
mix} = g_r + i g_i$ of the F\big|N interface. The thermally activated magnetization dynamics is
determined by the magnon temperature $T_F^m$, while the lattice and electron temperatures are
$T_F^p = T_F^e = T_F$. The term proportional to $g_r$ in $\II_{sp}$ has the same form as the
magnetic damping phenomenology of the Landau-Lifshitz Gilbert equation (introduced below in
\Eq{eqn:LLG}). The energy loss due to the spin current represented by $g_r$ therefore increases
the Gilbert damping constant. According to the Fluctuation-Dissipation Theorem (FDT), the noise
component of this spin-pumping induced current (from F to N) is accompanied by a fluctuating
spin current $\II_{fl}$ from the normal metal bath (from N to F). The latter is caused by the
thermal noise in N and its effect on F can be described by a random magnetic field $\hh'$
acting on the magnetization: \cite{foros_magnetization_2005}
\begin{equation}
	{\bf I}_{fl}(t) = - {M_sV\ov {\gamma}} {\gamma}~\mm(t){\times}\hh'(t).
\label{eqn:Ifl}
\end{equation}
In the classical limit (at high temperatures $k_BT{\gg} {\hbar}{\omega}_0$, where ${\omega}_0$ is the
ferromagnetic resonance frequency) $\hh'(t)$ satisfies the time correlation
\begin{equation}
	\avg{{\gamma}h'_i(t){\gamma}h'_j(0)} = {2{\alpha}'{\gamma}k_BT_N\ov M_sV}{\delta}_{ij}{\delta}(t)
	{\equiv}{{\sigma}'}^2{\delta}_{ij}{\delta}(t).
	\label{eqn:hphp}
\end{equation}
where $\Avg{{\cdots}}$ denotes the ensemble average, $i, j = x, y$, and ${\alpha}' = ({\gamma}{\hbar}/4{\pi}M_sV)g_r$
is the magnetization damping contribution caused by the spin pumping. The correlator is
proportional to the temperature $T_N$. 

The spin current flowing through the interface is given by the sum $\II_s = \II_{sp} +
\II_{fl}$ (see \Figure{fig:FN}). Here we are interested in the DC component:
\begin{equation}
	\avg{\II_s} =  {M_sV\ov {\gamma}} \midb{{\alpha}'~\avg{\mm{\times}\dmm} + {\gamma}~\avg{\mm{\times}\hh'}},
\end{equation}
At thermal equilibrium, $T_N = T_F^m$, and $\avg{\II_s} = 0$. At non-equilibrium situation, the
spin current component polarized along $\dmm$ with prefactor $g_i$ in \Eq{eqn:Isp} averages to
zero, thus does not cause observable effects on the DC properties in the present model. The
$\hxx$ and $\hyy$ component of $\avg{{\bf I}_s}$ also vanish and:
\begin{equation}
	\avg{I_z} = {M_sV\ov {\gamma}} 
	\midb{{\alpha}'\avg{m_x\dot{m}_y-m_y\dot{m}_x} - {\gamma}\avg{m_xh'_y-m_yh'_x}}.
	\label{eqn:Isz}
\end{equation}
We therefore have to evaluate the correlators: $\avg{m_i(0)\dm_j(0)}$ and
$\avg{m_i(0)h'_j(0)}$.

The motion of $\mm$ is governed by the Landau-Lifshitz-Gilbert (LLG) equation:
\begin{equation}
	\dot\mm = -{\gamma}~\mm{\times}\smlb{\Heff\hzz + \hh} + {\alpha}~\mm{\times}\dot\mm,
	\label{eqn:LLG}
\end{equation}
where $\Heff$ and ${\alpha}$ are the effective magnetic field and total magnetic damping,
respectively. $\hh$ accounts for the random fields associated with all sources of magnetic
damping, {\it viz.} thermal random field $\hh_0$ from the lattice associated with the bulk
damping ${\alpha}_0$, random field $\hh'$ from the N contact associated with enhanced damping ${\alpha}'$,
and possibly other random fields caused by, {\it e. g.} additional contacts. Random fields from
unrelated noise sources are statistically independent. The correlators of $\hh$ are therefore
additive and determined by the total magnetic damping ${\alpha} = {\alpha}_0 + {\alpha}' + {\cdots}$:
\begin{equation}
	\avg{{\gamma}h_i(t){\gamma}h_j(0)} = {2{\alpha}{\gamma}k_BT_F^m\ov M_sV}{\delta}_{ij}{\delta}(t) {\equiv} {\sigma}^2{\delta}_{ij}{\delta}(t).
	\label{eqn:hh}
\end{equation}
The magnon temperature $T_F^m$ is affected by the temperatures of and couplings to all
subsystems: ${\alpha} T_F^m = {\alpha}_0 T_F + {\alpha}' T_N + {\cdots}$.

We consider near-equilibrium situations, thus we may linearize the LLG equation. To first order
in $m_{\perp}{\ll}1$ (with $m_z{\simeq}1$)
\begin{subequations}
\label{eqn:LLG2}
\begin{align}
	\dot{m}_x + {\alpha}\dot{m}_y &= -{\omega}_0m_y + {\gamma}h_y, \\
	\dot{m}_y - {\alpha}\dot{m}_x &= +{\omega}_0m_x - {\gamma}h_x,
\end{align}
\end{subequations}
where ${\omega}_0 = {\gamma}\Heff$ is the ferromagnetic resonance (FR) frequency. With the Fourier
transform into frequency space $\td{g}={\int}g e^{i{\omega}t} dt$ and the inverse transform $g = {\int}
\td{g} e^{-i{\omega}t} d{\omega}/2{\pi}$, \Eq{eqn:LLG2} reads $\td{m}_i({\omega}) = {\sum}_j {\chi}_{ij}({\omega}) {\gamma}
\td{h}_j({\omega})$, with $i, j = x, y$ and the transverse dynamic magnetic susceptibility
\begin{equation}
	{\chi}({\omega}) = {1\ov ({\omega}_0-i{\alpha}{\omega})^2-{\omega}^2}
	\smlb{\begin{matrix} {\omega}_0-i{\alpha}{\omega}& -i{\omega}\\ i{\omega}& {\omega}_0-i{\alpha}{\omega}\end{matrix}},
	\label{eqn:chi}
\end{equation}
in terms of which, utilizing \Eqs{eqn:hphp}{eqn:hh},
\begin{subequations}
\begin{align}
	\Avg{m_i(t)m_j(0)} 
	&= {{\sigma}^2\ov 2{\alpha}} {\int}{{\chi}_{ij}({\omega})-{\chi}_{ji}^*({\omega})\ov i{\omega}}e^{-i{\omega}t} {d{\omega}\ov2{\pi}} , 
	\label{eqn:mm} \\
	\Avg{m_i(t)h'_j(0)} &= {{\sigma}'^2\ov{\gamma}} {\int}{\chi}_{ij}({\omega})e^{-i{\omega}t} {d{\omega}\ov2{\pi}}.
	\label{eqn:mh}
\end{align}
\end{subequations}
\Eq{eqn:mm} gives the mean square deviation of $\mm$ in the equal time limit $t\ra 0$:
$\avg{m_i(0)m_j(0)} = {\delta}_{ij}{\gamma}k_BT_F^m/{\omega}_0M_sV$. The time derivative of \Eq{eqn:mm} is
\begin{equation}
	\Avg{\dm_i(t)m_j(0)} 
	= -{{\sigma}^2\ov 2{\alpha}} {\int} \midb{{\chi}_{ij}({\omega})-{\chi}_{ji}^*({\omega})}e^{-i{\omega}t} {d{\omega}\ov2{\pi}},
	\label{eqn:mdm}
\end{equation}
By inserting \Eqs{eqn:mh}{eqn:mdm} with $t\ra 0$ into \Eq{eqn:Isz}:
\begin{align}
	\avg{I_z} 
	&= {M_sV\ov {\gamma}}
	\smlb{{{\alpha}'\ov {\alpha}}{\sigma}^2-{{\sigma}'}^2} {\int} \midb{{\chi}_{xy}({\omega})-{\chi}_{yx}({\omega})}{d{\omega}\ov2{\pi}} \nn
	& = {2{\alpha}'k_B\ov 1+{\alpha}^2} (T_F^m - T_N) 
	{\simeq} {{\gamma}{\hbar}g_r k_B\ov 2{\pi} M_sV}(T_F^m - T_N) \nn
	& {\equiv} L'_s(T_F^m - T_N),
	\label{eqn:Isz2}
\end{align}
where $L'_s$ is an interfacial spin Seebeck coefficient. In \Eq{eqn:Isz2}, we used ${\int}
\imm{{\chi}_{ij}({\omega})}d{\omega}/2{\pi} = 0$ (since \Eq{eqn:mh} is real and Im${\chi}$ changes sign when
${\omega}\ra-{\omega}$), and ${\int} \midb{{\chi}_{xy}({\omega}) - {\chi}_{yx}({\omega})}(d{\omega}/2{\pi}) = 1/(1+{\alpha}^2)$ (see Appendix
\ref{app:int}). From \Eq{eqn:Isz2} we conclude that the DC spin pumping current is proportional
to the temperature difference between the magnon and electron/lattice temperatures and
polarized along the average magnetization.

When the magnon temperature is higher (lower) than the lattice temperature, the DC spin pumping
current flows from F into N leading to a loss (gain) of angular momentum that is accompanied by
the heat current:
\begin{equation}
	Q_m = {2{\mu}_B\ov{\hbar}}\avg{I_z} \Heff = K'_mA(T_F^m - T_N),
	\label{eqn:Km}
\end{equation}
where $A$ is the contact area and $K'_m = {\omega}_0{\mu}_Bk_B(g_r/A)/{\pi}M_sV$ is the interface magnetic
heat conductance with Bohr magneton ${\mu}_B$.

In addition to the DC component of the spin pumping current, there is also an AC contribution
to the frequency power spectrum of spin current and spin Hall signal. \cite{xiao_charge_2009} A
measurement of the noise power spectrum should be interesting for insulating ferromagnets for
which the large imaginary part of the mixing conductance can be much larger than the real part
(see Appendix \ref{app:gmix}).

\subsection{Magnons} \label{sub:large}

In extended ferromagnetic layers the macrospin model breaks down and we have to consider magnon
excitations at all wave vectors. The space-time magnetization autocorrelation function can be
derived from the LLG equation (see Appendix \ref{app:mm}):
\begin{equation}
	\avg{\dm_i(0,0)m_i(0,0)} 
	= {{\gamma}k_BT_F^m \ov M_s V_a},
\end{equation}
where we introduced the temperature-dependent magnetic coherence volume
\begin{equation}
	V_a = {2\ov 3Z_{5\ov 2}}\smlb{4{\pi}D\ov k_BT_F^m}^{\hf{3}}
\label{eqn:Va}
\end{equation}
with $D$ the spin stiffness, and $Z$ the Zeta function. Physically, this coherence volume, or
its cube root, the coherence length, reflects the finite stiffness of the magnetic systems that
limits the range at which a given perturbation is felt. When this length is small, a random
field has a larger effect on a smaller magnetic volume.

The results obtained in Subsection II.A can be carried over simply by replacing $V\ra V_a$ in
\Eqss{eqn:hphp}{eqn:hh}. The corresponding spin Seebeck coefficient is $L_s' = {\gamma}{\hbar}k_B(g_r/A)/
2{\pi} M_s V_a$, and the heat conductance is $K'_m = {\omega}_0{\mu}_Bk_B(g_r/A)/{\pi}M_sV_a$. Here we assumed
that the magnon temperature does not change appreciably in the volume $V_a{\ll} V$.

\section{Magnon-phonon temperature difference profile}
\label{sec:TmTp}

Sanders and Walton (SW) \cite{sanders_effect_1977} discussed a scenario in which a
magnon-phonon temperature difference arises when a constant heat flow (or a temperature
gradient) is applied over an F insulator with special attention to the ferrimagnet yttrium iron
garnet (YIG). Its antiferromagnetic component is small and will be disregarded in the
following. This material is especially interesting because of its small Gilbert damping, which
translates into a long length scale of persistence of a non-equilibrium state between the
magnetic and lattice systems. SW assumes that, at the boundaries heat can only penetrate
through the phonon subsystem, whereas magnons can not communicate with the non-magnetic heat
baths ({\it i.e.} $K'_m = 0$ in our notation). Inside F bulk magnons interact with phonons and
become gradually thermalized with increasing distance from the interface. The different
boundary conditions for phonons and magnons lead to different phonon and magnon temperature
profiles within F. However, according to \Eq{eqn:Km}, the magnons in F are not completely
insulated as assumed by SW when the hear reservoirs are normal metals. In this section, we
follow SW and calculate the phonon-magnon temperature difference in an F insulator film induced
by the temperature bias but consider also the magnon thermal conductivity $K'_m$ of the
interfaces. In case of a metallic ferromagnet the conduction electron system provides an
additional parallel channel for the heat current.

As argued above, the boundary conditions employed by SW need to be modified when the heat baths
connected to the ferromagnet are metals. In that case, the magnons are not fully confined to
the ferromagnet, since a spin current can be pumped into or extracted out of the normal metal.
In \Figure{fig:NFNC}, the two ends of F are at different temperatures: $T_L$ and $T_R$ of the N
contacts drive the heat flow. Let us ignore the Pt contact on top of the F for now. When both
phonons and magnons are in contact with the reservoirs, by energy conservation (similar to Ref.
\onlinecite{sanders_effect_1977}) the integrated heat $Q_{mp}$ ($\bar{Q}_{mp}$) flowing from
the phonon to the magnon subsystem in the range of $-L/2\le z'\le z$ ($z\le z'\le L/2$) has the
following form:
\begin{subequations}
\label{eqn:TmTp}
\begin{align}
	Q_{mp}(z) &= {C_pC_m\ov C_T}{1\ov{\tau}_{mp}}{\int}_{-\hf{L}}^z[T_p(z')-T_m(z')]dz' \\
	          &= +K_m{dT_m\ov dz} + K'_m[T_m(-L/2) - T_L] \\
	          &= -K_p{dT_p\ov dz} - K'_p[T_p(-L/2) - T_L], \\
	\bar{Q}_{mp}(z) &= {C_pC_m\ov C_T}{1\ov{\tau}_{mp}}{\int}_z^{\hf{L}}[T_p(z')-T_m(z')]dz' \\
	                &= -K_m{dT_m\ov dz} + K'_m[T_m(L/2) - T_R] \\
	                &= +K_p{dT_p\ov dz} - K'_p[T_p(L/2) - T_R],
\end{align}
\end{subequations}
where $C_{p, m}$ ($C_T = C_p+C_m$) are the specific heats, $K_{p, m}$ ($K_T = K_p+K_m$) are the
bulk thermal conductivities for the phonon and magnon subsystems, $K'_{p, m}$ ($K'_T =
K'_p+K'_m$) are the respective boundary thermal conductivities. ${\tau}_{mp}$ is the magnon-phonon
thermalization (or spin-lattice relaxation) time. \cite{kittel_relaxation_1953} The boundary
conditions for $T_m$ and $T_p$ are set by letting $z={\pm}L/2$ in \Eq{eqn:TmTp}, {\it i.e.}
\begin{equation}
\label{eqn:bc}
	\left.K_{m,p}{dT_{m,p}\ov dz}\right|_{z = {\pm}\hf{L}} 
	= {\mp}K'_{m,p}[T_{m,p}({\pm}L/2) - T_{R/L}], \\
\end{equation}
The solution to \Eq{eqn:TmTp} and \Eq{eqn:bc} yields the magnon-phonon temperature difference
${\Delta}T_{mp}(z) = T_m(z) - T_p(z)$:
\begin{widetext}
\begin{equation}
	{\Delta}T_{mp}(z)
	= {K_T(K'_pK_m-K'_mK_p)\sinh{z\ov{\lambda}}~{\Delta} T \ov 
	{1\ov{\lambda}}K_mK_p(K'_TL+2K_T) \cosh{{L\ov2{\lambda}}}
	+ (2K'_mK_p^2+2K'_pK_m^2+K'_pK'_mLK_T) \sinh{{L\ov2{\lambda}}} }
	{\equiv} {\eta}~{\sinh {z\ov{\lambda}}\ov \sinh{{L\ov2{\lambda}}}}~{\Delta} T
	\label{eqn:dTmp}
\end{equation}
\end{widetext}
with ${\Delta}T = T_L - T_R$ and
\begin{equation}
	{\lambda}^2 = {C_p+C_m\ov C_pC_m}{K_pK_m\ov K_p+K_m}{\tau}_{mp} 
	{\approx} {K_m\ov C_m}{\tau}_{mp},
\label{eqn:lambda0}
\end{equation}
where the approximation applies when $C_p{\gg} C_m$ and $K_p{\gg} K_m$.
\Eq{eqn:dTmp} shows that the deviation of the magnon temperature from the lattice (phonon)
temperature is proportional to the applied temperature bias and decays to zero far from the
boundaries with characteristic (magnon diffusion) length ${\lambda}$.

We use the diffusion limited magnon thermal conductivity and specific heat calculated by a
simple kinetic theory (assuming ${\hbar}{\omega}_0{\ll} k_BT$): \cite{yelon_magnon_1972}
\begin{equation*}
	C_m = {15{\zeta}({5\ov2})\ov 32}\sqrt{k_B^5T^3\ov {\pi}^3D^3} \qand
	K_m = {35{\zeta}({7\ov2})\ov 16}\sqrt{k_B^7T^5\ov {\pi}^3D}{{\tau}_m\ov {\hbar}^2}
\label{eqn:KmCm}
\end{equation*}
with ${\tau}_m$ the magnon scattering time. Using these expressions in the approximate form of
\Eq{eqn:lambda0} we obtain:
\begin{equation}
	{\lambda}^2
	= {14 {\zeta}(7/2)\ov 3{\zeta}(5/2)}{D k_BT\ov{\hbar}^2}{\tau}_m{\tau}_{mp}.
	\label{eqn:lambda}
\end{equation}
In Appendix \ref{app:atmp}, we estimate ${\tau}_{mp}{\simeq}1/(2{\alpha}{\omega}_0)$ for ferromagnetic insulators,
assuming that magnetic damping ${\alpha}$ is caused by magnon-phonon scattering. It is difficult to
estimate or measure ${\tau}_m$, and the values quoted in the literatures ranges from $10^{-9}$s to
$10^{-7}$s, depending on both material and temperature. \cite{kittel_relaxation_1953,
demokritov_bose-einstein_2006} At present we cannot predict how ${\lambda}$ varies with temperature:
\Eq{eqn:lambda} seems to increase with temperature, but the relaxation times likely decrease
with T.

When $K'_m \ra 0$ and $K'_p \ra {\infty}$, {\it i.e.} the boundary is thermally insulated for
magnons and has zero thermal resistivity for phonons, the prefactor ${\eta}$ in \Eq{eqn:dTmp}
reduces to SW's result: \cite{sanders_effect_1977}
\begin{equation}
	{\eta} = {K_T \ov {L\ov{\lambda}}K_p \coth{L\ov2{\lambda}}+ 2K_m } {\approx} {1\ov {L\ov{\lambda}} \coth{L\ov2{\lambda}}},
	\label{eqn:eta}
\end{equation}
where the approximation is valid when $K_p{\gg} K_m$. ${\Delta}T_{mp}$ is obviously maximal in this
limit.

The discussion in this section also applies to ferromagnetic metals when the electron-phonon
relaxation is much faster than the magnon-phonon relaxation. The electron and phonon subsystem
are then thermalized with each other and can be treated as one subsystem. In this case, we may
replace $K_p \ra K_{pe} = K_p + K_e$ and
\begin{equation}
	{C_pC_m\ov C_p+C_m}{1\ov {\tau}_{mp}} \ra 
	{C_pC_m\ov C_p+C_m}{1\ov {\tau}_{mp}} +
	{C_eC_m\ov C_e+C_m}{1\ov {\tau}_{me}}.
\label{eqn:rep}
\end{equation}

\section{Hall voltage}
\label{sec:VH}

\begin{table}[t]
	\centering
	\begin{tabular}{l|r@{}l|r@{}l|l} \hline
					& 		& YIG			& 		&Py				& Unit \hhline
		${\gamma}$ 		& 		&$1.76{\times}10^{11}$ & 		&$1.76{\times}10^{11}$	& 1/T${\cdot}$s \\ 
		$4{\pi}M_s$		& $^a$	&$1.4{\times}10^5$		& $^f$	&$8.0{\times}10^5$ 	& A/m \\ 
		$D$			& $^b$	&$1.55{\times}10^{-38}$	& $^f$	&$7.6{\times}10^{-39}$	& J${\cdot}$m$^2$ \\ 
		${\alpha}$			& $^a$	&$5{\times}10^{-5}$	& $^g$	&0.01			& --- \\ 
		${\omega}_0$		& $^a$	&10				& $^g$	&20				& GHz \\ 
		${\tau}_{mp}$	& $^{c,d}$&$10^{-6}$	& $^d$	&$10^{-7}$ (Ni)	& s \\ 
		${\tau}_m$		& $^{c,e}$&$10^{-9{\sim}7}$ 	& $^h$	&$10^{-9}$		& s \\ 
		${g_r\ov A}$& $^a$	&$10^{15{\sim}16}$ 		& $^i$	&$10^{18}$		& 1/m$^2$ \\
		$V_a^{1/3}$	& 		&5.4			& 		&3.8			& nm \\
		${\eta}$			& 		&0.4 - 0.5			& 		&0.27			& --- \\ \hline
		${\lambda}$ (th)	& 		&4.7 - 47		& 		&0.3			& mm \\
		${\lambda}$ (exp)	& $^a$	&6.7			& $^k$	&4.0			& mm \\
		${\xi}$ (th) 	& 		&0.38 - 3.8			& 		&130			& ${\mu}$V/K \\
		${\xi}$ (exp)	& $^a$	&0.16			& $^k$	&0.25			& ${\mu}$V/K \\ \hline 
	\end{tabular}
	\caption{Parameters for YIG and Py (at $T = 300$K if not specified, LT ${\approx}$ 10 K).
		$^a$Ref. \onlinecite{kajiwara_transmission_2010},
		$^b$Ref. \onlinecite{uchida_unpublished},
		$^c$Ref. \onlinecite{demokritov_bose-einstein_2006},
		$^d$Ref. \onlinecite{spencer_spin_1960}, 
		$^e$Ref. \onlinecite{kittel_relaxation_1953},
		$^f$Ref. \onlinecite{roy_antivortex_2009} ($D$ is derived from $A_{\rm ex} = 13$ pJ/m),
		$^g$Ref. \onlinecite{vlaminck_current-induced_2008},
		$^h$Ref. \onlinecite{hsu_magnon_1976},
		$^i$Ref. \onlinecite{brataas_non-collinear_2006},
		$^j$Ref. \onlinecite{uchida_observation_2008}.
		}
	\label{tab:param}
\end{table}

\begin{table}
	\centering
	\begin{tabular}{r|r|c}
		\hline
		Quantity 	& Values 				& Reference \hhline
		${\theta}_H$		& 0.0037 				&  \onlinecite{kimura_room-temperature_2007}\\
		${\rho}$			& 0.91 ${\mu}{\Omega}{\cdot}$m 			& \onlinecite{uchida_observation_2008} \\
		$l{\times}w{\times}h$		& 4 mm ${\times}$ 0.1 mm ${\times}$ 15 nm & \onlinecite{uchida_observation_2008} \\ \hline
	\end{tabular}
	\caption{Parameters for Pt contact.}
	\label{tab:Pt}
\end{table}

In Uchida {\it et al.}'s experiment (Ref. \onlinecite{uchida_observation_2008}), a Pt contact
is attached on top of a Py film (see \Figure{fig:NFNC}) to detect the spin current signals by
the inverse spin Hall effect. A spin current polarized in the $\hzz$-direction that flows into
the contact in the $\hyy$-direction is converted into an electric Hall current $I_c$ and thus a
Hall voltage $V_H = I_c R$ in the $\hxx$-direction.

For the setup shown in \Figure{fig:NFNC} we assume that the Pt contact is small enough to not
disturb the system, which is valid when the heat flowing into Pt is much less than the heat
exchange between the magnons and phonons or
\begin{equation*}
	K'_m{\Delta}T_{mp}(z)(l{\times}w) {\ll}  {C_pC_m\ov C_T}{1\ov {\tau}_{mp}}{\Delta}T_{mp}(z)(l{\times}w{\times}d),
\end{equation*}
which is well satisfied when $d{\gg} 1$ nm for both YIG and Py at low temperatures.

From \Eq{eqn:dTmp}, we see that at the position below the contact ($z = z_c$) the magnon
temperature deviates from the lattice (and electron) temperatures by $T_F^m - T_F^p =
{\Delta}T_{mp}(z_c)$. If we assume that the contact is at thermal equilibrium with the lattice (and
electrons) in the F film underneath, {\it i.e.} $T_c^e = T_c^p = T_F^p(z_c)$, then a
temperature difference between the magnons in F and the electrons in Pt exists: $T_F^m(z_c) -
T_c^e = {\Delta}T_{mp}(z_c)$. Therefore, by \Eq{eqn:Isz2}, a DC spin-pumping current $\avg{I_z}$ from
F to Pt is driven by this temperature difference, which gives rise to a DC Hall current in the
Pt contact (see \Figure{fig:NFNC}):
\begin{equation}
	j_c \hxx = {\theta}_H {2e\ov{\hbar}} {\avg{I_z}\ov A} \hzz{\times}\hyy,
	\label{eqn:Ic}
\end{equation}
where ${\theta}_H$ is the Hall angle. In \Eq{eqn:Ic} we disregarded the spin diffusion backflow and
finite thickness corrections to the Hall effect, which is reasonable when thickness $h$ is
comparable to the spin diffusion length $l_{sd}$.

According to \Eq{eqn:Isz2}, the thermally driven spin pumping current can flow from F to Pt:
\begin{equation}
	\avg{I_z} = L'_s[T_m(z_c) - T_c] = L'_s[T_m(z_c) - T_p(z_c)].
\end{equation}
Making use of \Eq{eqn:dTmp}, the electric voltage over the two transverse ends of the Pt
contact separated by distance $l$ is $V_H(z_c) = {\rho} l j_c(z_c)$:
\begin{equation}
	V_H
    = {{\eta}{\theta}_H{\rho} l g_r {\gamma}ek_B\ov {\pi} M_s V_a A} {\sinh{z_c\ov{\lambda}}\ov\sinh{{L\ov2{\lambda}}}} {\Delta}T 
	{\equiv} {\xi}~{\sinh{z_c\ov{\lambda}}\ov\sinh{{L\ov2{\lambda}}}} {\Delta}T,
	\label{eqn:VH}
\end{equation}

Using the numbers in \Table{tab:param} and \Table{tab:Pt} and a value of ${\lambda} {\simeq} 7$ mm that
reflects the low magnetization damping in YIG, we have ${\eta} {\simeq} 0.47$ from \Eq{eqn:eta}, and ${\xi}
{\simeq} 0.38{\sim}3.8{\mu}$V/K from \Eq{eqn:VH} for YIG for $g_r/A {\simeq} 10^{15{\sim}16}$/m$^2$, which is
consistent with with the experiments by Uchida {\it et al.} (Ref.
\onlinecite{uchida_unpublished}). For Py, we estimate ${\xi} {\simeq} 130~{\mu}$V/K, which is almost three
orders of magnitude larger than the experimental value of $0.25~{\mu}$V/K given in Ref.
\onlinecite{uchida_observation_2008}.

The crucial length scale ${\lambda}$ is determined by two thermalization times: the magnon-magnon
thermalization time ${\tau}_m$ and the magnon-phonon thermalization time ${\tau}_{mp}$ as shown in
\Eq{eqn:lambda}. Knowledge of these two times is essential in estimating ${\lambda}$. The quoted
values of ${\tau}_{m, mp}$ are rough order of magnitude estimates at low temperatures. Yet, in
order to completely pin down the value of ${\lambda}$ for this material, a more accurate determination
of ${\tau}_{m, mp}$ and $K_m, C_m$ as a function of temperature by both theory and experiments is
required. As seen in \Eq{eqn:VH}, the spatial variation of the Hall voltage over the F strip is
determined by the magnon-phonon temperature difference profile as calculated in the previous
section. From \Eq{eqn:lambda} and the parameters for YIG in \Table{tab:param} (where the values
of ${\tau}_m$ and ${\tau}_{mp}$ are very uncertain), we estimate ${\lambda} {\simeq} 4.7 {\sim} 47$ mm, which is again
consistent with the experimental value of ${\lambda} {\simeq} 6.7$ mm. For Py, we estimate ${\lambda} {\simeq} 0.3$ mm
(using the ${\tau}_{mp}$ value for Ni instead of Py), which is about one order of magnitude smaller
than its measured value.

\section{Discussion \& Conclusion}
\label{app:disc}

A magnon-phonon temperature difference drives a DC spin current, which can be detected by the
inverse spin Hall effect or other techniques. This effect can, {\it vice versa}, be used to
measure the magnon-phonon temperature difference. Because the spatial dependence of this effect
relies on various thermalization times between quasi-particles, which are usually difficult to
measure or calculate, this effect also accesses these thermalization times. On a basic level,
we predict that the spin Seebeck effect is caused by the non-equilibrium between magnon and
phonon systems that is excited by a temperature basis over a ferromagnet. Since the inverse
spin Hall effect only provides indirect evidence, it would be interesting to measure the
presumed magnon-phonon temperature difference profile by other means. Such measurements would
also give insight into relaxation times that are difficult to obtain otherwise.

In principle, the theory holds for both ferromagnetic insulators and metals. However, as shown
above, the agreement between the theory and experiments is reasonably good for ferromagnetic
insulator YIG, but the theory fails for ferromagnetic metal Py, which underestimates the length
scale ${\lambda}$ and overestimates the magnitude ${\xi}$. This might be because of several reasons: (i)
we have completely ignored the short-circuiting effect of the metallic Py, to which the inverse
spin Hall current leaks, (ii) the lack of reliable information about relaxation times ${\tau}_{mp,
m}$ for Py ccould cause the difference in ${\lambda}$, (iii) the complication due to the existence of
conduction electrons in ferromagnetic metals.

In conclusion, we propose a mechanism for the spin Seebeck effect based on the combination of:
(i) the inverse spin Hall effect, which converts the spin current into an electrical voltage,
(ii) thermally activated spin pumping at the F\big|N interface driven by the phonon-magnon
temperature difference, and (iii) the phonon-magnon temperature difference profile induced by
the temperature bias applied over a ferromagnetic film. Effect (ii) also introduces an
additional magnon contributed thermal conductivity of F\big|N interfaces. The theory holds for
both ferromagnetic metals and insulators. The agreement between experiments and theory is
satisfactory for insulating ferromagnet. The magnitude and the spatial length scale for YIG is
predicted to be in microvolt and millimeter range. The lack of agreement for both the length
scale and the magnitude of the spin Seebeck effect for permalloy remains to be explained.

\section*{Acknowledgment}

This work has been supported by EC Contract IST-033749 ``DynaMax'', a Grant for Industrial
Technology Research from NEDO, Japan, and National Natural Science Foundation of China (Grand
No. 10944004). J. X. and G. B. acknowledge the hospitality of the Maekawa Group at the IMR,
Sendai.

\appendix

\section{Integral} \label{app:int}

Here we evaluate the frequency integral in \Eq{eqn:Isz2}. To this end, we need to reintroduce
the time dependence and then take the limit $t\ra 0$. When $t>0$, the integral in \Eq{eqn:mh}
can be calculated using contour integration (real axis + semi-circle in the lower plane so that
$e^{-i{\omega}t}\ra 0$, where the integral over the semi-circle vanishes by {\it Jordan's Lemma}):
\begin{equation}
	{\int}{\chi}e^{-i{\omega}t}{d{\omega}\ov2{\pi}} 
	= -i~\imm{ {e^{-i{\omega}_-t}\ov 1-i{\alpha}} (\hunit-\hsigma_y)}{\Theta}(t). 
\end{equation}
The minus sign comes from the counter-clockwise contour, and ${\Theta}(t)$ is the Heaviside step
function, which vanishes when $t<0$ and is unity otherwise. This integral is discontinuous at
$t = 0$, therefore the value at $t=0$ is given by the average of the values at $t = 0^{\pm}$:
\begin{equation}
	{\int}{\chi}e^{-i{\omega}0}{d{\omega}\ov2{\pi}}
	= \half{1\ov 1+{\alpha}^2} \smlb{\begin{matrix}  {\alpha}& 1 \\ -1 & {\alpha}\end{matrix}}.
\label{eqn:intchi}
\end{equation}

\section{Spin-mixing conductance for F\big|N interface} \label{app:gmix}

\begin{figure}[t]
	\includegraphics[width=0.95\columnwidth]{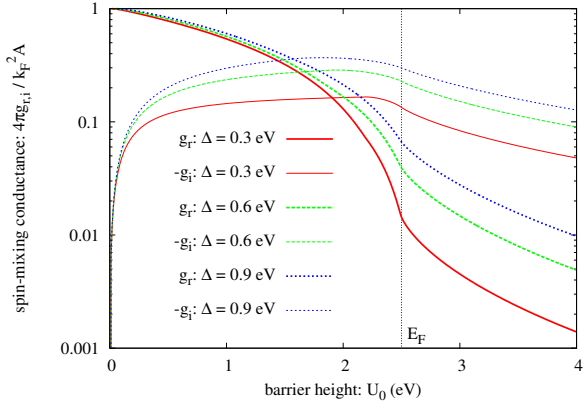}
	\caption{(Color online) $g_{r,i}$ vs. $U_0$ in logscale for different ${\Delta}$ values. Curves
	starting with values close to unity give the real part $g_r$, and curves starting from
	very low values display the imaginary part $-g_i$.  The vertical axis is in units of the
	Sharvin conductance.}
	\label{fig:gmix}
\end{figure}

In this Appendix, we use a simple parabolic band model to estimate the spin pumping at an
F\big|N interface, where F can be a conductor or an insulator. The Fermi energy in N is $E_F$,
and the bottom of the conduction band for spin up and spin down electrons in F are at $E_F +
U_0$ and $E_F + U_0 + {\Delta}$, respectively, where $U_0$ is the bottom of the majority band and
${\Delta}$ is the exchange splitting. The reflection coefficient for an electron spin ${\sigma}$ ($\up$ or
$\dn$) from N at the Fermi energy reads
\begin{equation}
	r_{\sigma}(k) = {k - k_{\sigma}\ov k + k_{\sigma}}
\label{eqn:r}
\end{equation}
where $k = \sqrt{2m_0E_F/{\hbar}^2 - q^2}$ is the longitudinal wave-vector in N ($q$ is the
transverse wave-vector). $k_\up = \sqrt{2m_\up (E_F - U_0)/{\hbar}^2 - q^2}$ and $k_\dn =
\sqrt{2m_\dn (E_F - U_0 - {\Delta})/{\hbar}^2 - q^2}$ are the longitudinal wave-vectors (or imaginary
decay constants) in F for both spins. $m_0$ is the effective mass in N and $m_{\sigma}$ is the
effective mass for spin ${\sigma}$ in F. The mixing conductance reads
\begin{equation}
	g_{\rm mix} 
	= {A\ov 4{\pi}^2}{\int} \smlb{1 - r_\up r_\dn^*} d^2{\bf q}
	= g_r + i g_i.
\label{eqn:gmix}
\end{equation}
We evaluate \Eq{eqn:gmix} with $m_0 = m_{\sigma}$ having the free electron mass, $E_F = 2.5$ eV, $U_0
= 0 - 4$ eV. \cite{jin_barrier-height_2006} A plot of the mixing conductance is shown in
\Figure{fig:gmix} for ${\Delta} = 0.3, 0.6, 0.9$ eV. $U_0 = 0$ corresponds to a ferromagnetic metal
(with majority band matching the N electronic structure), and $U_0 \ge E_F = 2.5$ eV
corresponds to a ferromagnetic insulator (with $\abs{r_{\sigma}} = 1$ for both spin types).

\section{Magnetization correlation for macroscopic samples} \label{app:mm}

The dynamics of $\mm$ is governed by the LLG equation:
\begin{equation}
	\dmm(\rr,t) = -{\gamma}\mm(\rr,t){\times}\HHeff + {\alpha}\mm(\rr,t){\times}\dmm(\rr,t),
	\label{eqn:LLG0}
\end{equation}
where $\HHeff = \HH_0 + (D/{\gamma}{\hbar}){\nabla}^2\mm(\rr,t) + \hh(\rr,t)$ is the total effective magnetic
field with: (i) the external field plus the uniaxial anisotropy field $\HH_0 = H_0\hzz =
({\omega}_0/{\gamma}) \hzz$, (ii) the exchange field $(D/{\gamma}{\hbar}){\nabla}^2\mm$ due to spatial variation of
magnetization, and (iii) the thermal random fields $\hh(\rr,t)$.

We define Fourier transforms:
\begin{align}
	\td{g}(\kk,{\omega}) &= {\int}d\rr e^{i\kk{\cdot}\rr}{\int}dt e^{i{\omega}t} ~g(\rr,t) \\
	g(\rr,t) &= {1\ov V}{\sum}_{m,n,l}e^{-i\kk{\cdot}\rr}{\int} {d{\omega}\ov 2{\pi}} e^{-i{\omega}t}~\td{g}(\kk,{\omega}) 
\label{eqn:FT}
\end{align}
where $\kk=2{\pi}(m/L, n/W, l/H)$ and $L, W, H$ are the length, width, and height of the
ferromagnetic film. Near thermal equilibrium, $m_{\perp}{\ll}1$ (and so $m_z{\simeq}1$), we may linearize
the LLG equation. After Fourier transformation, the linearized LLG equation becomes
$\td{m}_i(\kk,{\omega}) = {\sum}_j {\chi}_{ij}(\kk,{\omega}){\gamma}\td{h}_j(\kk, {\omega})$ with $i, j = x, y$, and
\begin{equation} 
	{\chi}
	= -{1/(1+{\alpha}^2)\ov ({\omega}-{\omega}_\kk^+)({\omega}-{\omega}_\kk^-)} 
	\smlb{\begin{matrix} {\omega}_\kk-i{\alpha}{\omega}& -i{\omega} \\ i{\omega}& {\omega}_\kk-i{\alpha}{\omega}\end{matrix}}
	\label{eqn:chikw}
\end{equation}
with ${\hbar}{\omega}_\kk = {\hbar}{\omega}_0+D\abs{\kk}^2$ and ${\omega}_\kk^{\pm} = {\pm}{\omega}_\kk/(1{\pm}i{\alpha})$.

The spectrum of random motion in the linear response regime $x_i = {\sum}_j {\chi}_{ij} f_j$ with
``current'' $x$ and random force $f$ ($x$ and $f$ are chosen such that $xf$ is in the units of
energy) is comprehensively studied in Ref. \onlinecite{landau_statistical_1980}. In our
problem, $f\ra {\gamma}h$, $x\ra (M_s/{\gamma})m$, ${\chi}\ra(M_s/{\gamma}){\chi}$, the autocorrelation of the
magnetization then becomes ($\bm{\zeta} = \rr_1-\rr_2, {\tau} = t_1 - t_2$)
\begin{align}
	&\Avg{m_i(\bm{{\zeta}}, {\tau})m_j(0, 0)}
	= {2{\gamma}k_BT\ov M_sV} \nn
	&{\times} {\sum}_\kk e^{-i\kk{\cdot}\bm{{\zeta}}}{\int}{d{\omega}\ov2{\pi}}e^{-i{\omega}{\tau}}
	{{\chi}_{ij} - {\chi}_{ji}^*\ov i{\omega}} {{\hbar}\abs{{\omega}}/k_BT\ov e^{{\hbar}\abs{{\omega}}/\kB T}-1},
	\label{eqn:mm120}
\end{align}
from which $\Avg{\dm_i(\bm{{\zeta}}, {\tau})m_j(0, 0)}$ can be caluclated by taking derivative over
${\tau}$.

The limit $\bm{{\zeta}} = 0$ and ${\tau} = 0$ can be obtained for three-dimensional systems by replacing
the sum over $\kk$ by an integral $V^{-1}{\sum}_\kk\ra{\int}d^3\kk/(2{\pi})^3$ in \Eq{eqn:mm120}:
\begin{align}
	\Avg{m_i(0, 0)m_j(0, 0)}
	&= Z_{3\ov 2}{{\gamma}{\hbar}\ov M_s} \smlb{\kB T\ov 4{\pi}D}^\hf{3} {\delta}_{ij}, \\
	\Avg{\dm_i(0, 0)m_j(0, 0)}
	&= {3Z_{5\ov 2}{\gamma}k_BT\ov 2M_s} \smlb{\kB T\ov 4{\pi}D}^\hf{3}
	\smlb{\begin{matrix}  -{\alpha} & -1 \\ 1 & -{\alpha} \end{matrix}}_{ij},
\label{eqn:C003D}
\end{align}
where $Z$ is the Zeta function.


\section{Relation between ${\alpha}$ and ${\tau}_{mp}$} \label{app:atmp}

Here we derive a relationship between the magnetic damping constant ${\alpha}$ and the magnon-phonon
thermalization time ${\tau}_{mp}$. When the magnon and phonon temperatures are $T_m$ and $T_p$, the
magnon-phonon relaxation time is phenomenologically defined by: \cite{sanders_effect_1977}
\begin{equation}
	{d\ov dt}(T_p - T_m) = - {T_p - T_m \ov {\tau}_{mp}}.
\label{eqn:taump}
\end{equation}
If the phonon system is attached to a huge reservoir (substrate) such that its temperature
$T_p$ is fixed ($dT_p/dt = 0$), \Eq{eqn:taump} becomes
\begin{equation}
	{dT_m\ov dt} {\simeq} - {T_m - T_p \ov {\tau}_{mp}}.
\label{eqn:taump2}
\end{equation}
In metals we should consider three subsystems (magnon, phonon, electron) leading to
\begin{equation}
	{d T_s\ov dt} = - {\sum}_t k_{st} (T_s - T_t),
\end{equation}
with $s, t =$ m, p, e for magnon, phonon, and electron. The coupling strength $k_{st} =
{\tau}_{st}^{-1}C_t/(C_s + C_t)$ with the $s$-$t$ relaxation time ${\tau}_{st}$ and the specific heat
$C_s$. When the magnon specific heat is much less than that of phonons and electrons ($C_m {\ll}
C_p, C_e$)
\begin{equation}
	{dT_m\ov dt} {\simeq} -{T_m - T_p \ov {\tau}_{mp}} - {T_m - T_e\ov {\tau}_{me}}.
\label{eqn:TmTpTe}
\end{equation}

We may parameterize the magnon temperature by the thermal suppression of the average
magnetization:
\begin{equation}
	\Heff M_sV (1 - \avg{m_z}) = k_B T_m,
\label{eqn:Tm}
\end{equation}
where $\Heff$ points in the $\hzz$-direction. The LLG equation provides us with the information
about $m_z$:
\begin{equation}
	\dmm = - {\gamma} \mm{\times}(\HHeff + \hh) + {\alpha}~\mm{\times}\dmm,
\end{equation}
where the random thermal field $\hh = \hh_p + \hh_e$ from the lattice and electrons are
determined by the phonon/electron temperature:
\begin{equation}
	\avg{{\gamma}h_{p/e}^i(t){\gamma}h_{p/e}^j(0)} = {2{\gamma}{\alpha}_{p/e}k_B T_{p/e}\ov M_sV}
\end{equation}
with ${\alpha}_{p/e}$ the magnetic damping caused by scattering with phonons/electrons and ${\alpha} = {\alpha}_p
+ {\alpha}_e$. Therefore
\begin{widetext}
\begin{equation}
	- {k_B\ov \Heff M_sV} {dT_m\ov dt} 
	= \avg{\dm_z}
	= \avg{m_y{\gamma}h_x - m_x{\gamma}h_y} + {\alpha}\avg{m_x\dm_y - m_y\dm_x} 
	= {1\ov 1+{\alpha}^2}{2{\gamma}k_B({\alpha}T_m-{\alpha}_pT_p-{\alpha}_eT_e)\ov M_sV}, 
\end{equation}
\end{widetext}
or
\begin{equation}
	{dT_m\ov dt} = - {2{\omega}_0\ov 1+{\alpha}^2} ({\alpha}T_m-{\alpha}_pT_p-{\alpha}_eT_e),
\label{eqn:TmTpTe2}
\end{equation}
with ${\omega}_0 = {\gamma}\Heff$. Comparing \Eq{eqn:TmTpTe2} with \Eq{eqn:TmTpTe}, we find
\begin{equation}
	2{\alpha}{\omega}_0 {\simeq} {2{\alpha}{\omega}_0\ov 1+{\alpha}^2} = {2{\alpha}_p{\omega}_0\ov 1+{\alpha}^2} + {2{\alpha}_e{\omega}_0\ov 1+{\alpha}^2} 
	= {1\ov {\tau}_{mp}} + {1\ov {\tau}_{me}}.
\label{eqn:alphatau}
\end{equation}
For the ferromagnetic insulator YIG: ${\alpha} {\simeq} 6.7{\times}10^{-5}$ and ${\omega}_0 {\simeq} 10$ GHz (Ref.
\onlinecite{kajiwara_transmission_2010}), thus ${\tau}_{mp} = 1/2{\alpha}{\omega}_0 {\simeq} 10^{-6}$ s, which
agrees with ${\tau}_{mp}{\simeq} 10^{-6} $ s in Ref. \onlinecite{sanders_effect_1977}.

The estimate above relies on the macrospin approximation. Considering magnons from all $\kk$
adds a prefactor of order of unity.


\end{document}